# SportSense: Real-Time Detection of NFL Game Events from Twitter

Technical Report TR0511-2012

Siqi Zhao*   Lin Zhong*   Jehan Wickramasuriya§   Venu Vasudevan§   Robert LiKamWa*   Ahmad Rahmati*

*Department of Electrical & Computer Engineering, Rice University, Houston, Texas, USA
§Betaworks, Applied Research Center, Motorola Mobility, Libertyville, Illinois, USA

**ABSTRACT**

We report our experience in building a working system, SportSense, which exploits Twitter users as human sensors of the physical world to detect events in real-time. Using the US National Football League (NFL) games as a case study, we report in-depth measurement studies of the delay and post rate of tweets, and their dependence on other properties. We subsequently develop a novel event detection method based on these findings, and demonstrate that it can effectively and accurately extract game events using open access Twitter data. SportSense has been evolving during the 2010-11 and 2011-12 NFL seasons and is able to recognize NFL game big plays in 30 to 90 seconds with 98% true positive, and 9% false positive rates. Using a smart electronic TV program guide, we show that SportSense can utilize human sensors to empower novel services.

## 1. Introduction

Twitter has over 200 million users, who can be can be collectively regarded as human sensors that provide quick, brief output for whatever motivates them to tweet. Many have attempted to interpret the output of these human sensors, including efforts that detect major social and physical events such as earthquakes [1], celebrity deaths [2], and presidential elections [3]. However, these work have significant delays and usually perform their processing off-line.

In this work, we answer a complementary question: *how good are these human sensors for the real-time detection of less significant but more frequent events, such as those happening in a nationally televised sports game?* The insights gained from answering this question will fuel novel applications that leverage the human sensors in real-time. For example, TV program guide can be enhanced with real-time game information before ESPN discloses game scores, as we will show in Section 6. We envision other applications as well, such as reporting traffic accidents.

To answer the above question, we make three contributions: (*i*) We present a first-of-its-kind measurement study of the key sensing properties of human sensors, i.e., *delay* and *post rate*, and their dependency on other properties such as device being used, user activeness, and the length of tweets. We find that tweets from different user categories had different properties in terms of delay and post rate. (*ii*) We leverage these properties to devise a novel detection algorithm that employs matched filter detector and form different event templates for these three fundamental categories of tweets. (*iii*) Finally, we implement the solutions to improve an in-house built web service, http://sportsense.us, to analyze tweets collected during US National Football League (NFL) games to detect big plays, such as scoring events and turnovers, tens of seconds after they happen. SportSense has been continuously operational for the past two NFL seasons, providing game information to NFL fans through web browser as well as its API, usually faster than EPSN website updates.

While SportSense was built for and tested through NFL games, its techniques can be readily applied to other sports games and even TV shows that have a similarly sized fan population and similar frequencies of major events, e.g., soccer, baseball, and reality shows. The keyword-based tweet retrieval method adopted by SportSense allows it to be easily revised for detecting other events of the physical world by updating the keywords.

## 2. Related Work

To the best of our knowledge, the measurements of human sensor delay and its dependency on other properties are novel. There has been extensive work done on related topics, from topological properties, users' properties, to event and sentiment detection. None of the existing solutions can detect or aim at detecting events in seconds. Most of the solutions are intended to be used off-line, instead of in real-time. Many of the existing solutions treat tweets as text documents and apply text mining techniques to extract topics, classify and cluster tweets, e.g., TwitterStand [2], Tweet the debates [3], and Topical clustering of tweets [4]. Some have extracted information about social and physical events using Twitter or similar services [1, 5-11]. A particularly relevant work, TwitInfo [12], presented a system for visualizing and summarizing events on Twitter. None of the above work can detect targeted events in seconds despite the "real-time" claim by some. For example, TwitInfo has a delay in the magnitude of minutes. The solution in [1] detects an earthquake from Twitter in a matter of hours. In contrast, we are interested in detecting events in seconds; and SportSense is able to reliably detect touchdowns in a few tens of seconds, even before ESPN updates the score on the web page.

Event detection for sports games has been studied by the video analysis research community [6, 13-16]. Unfortunately, the video content and text information

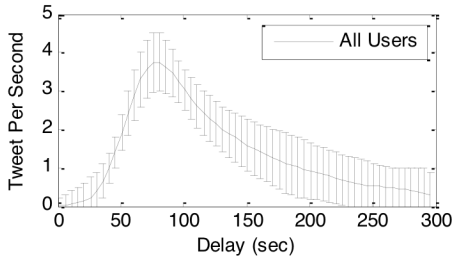

**Figure 1: The tweet response among all tweets.**

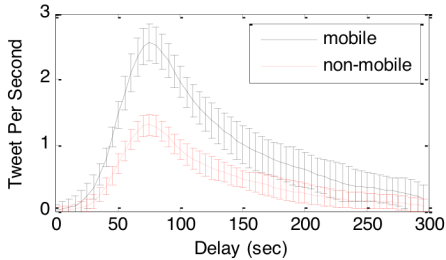

**Figure 2: The tweet response from mobile and non-mobile devices.**

leveraged by the above work is not always available, especially in real-time.

### 3. NFL Game Tweet Collecting and Event Tagging

During American football games, Twitter has a high volume of tweets from football fans, tweeting about football events that they find exciting or notable. SportSense leverages this activity, associating particular streams of tweets with game events (e.g. Touchdowns, Field Goals) to perform robust event detection.

#### 3.1 Collecting Related Twitter Data

Twitter provides various Application Programming Interfaces (APIs) for third parties to retrieve tweets from the general public. SportSense uses the free Streaming API [17], which provides a continuous flow of tweets in real-time. This API returns public tweets that match one or more filter predicates (e.g. keyword, username, and location). The latency from tweet creation to delivery on the API is typically within one second [18].

The Streaming API is subject to a few limitations. According to Twitter, excessive connection attempts result in an IP address ban. Additionally, broad predicates retrieve a sampled stream with only ~1% sample rate [19]. During the 2011 and 2012 Super Bowl, we discovered an undocumented limit of about 50 tweets per second [20].

Therefore, when analyzing tweets in real-time for a topic of interest, it is of utmost importance to focus on only relevant tweets. Irrelevant tweets not only take up precious collection space, but also incur computation cost in our analysis, where speed is critical to achieve real-time performance. To avoid these issues, SportSense employs team names and game event keywords to suppress irrelevant tweets. We found this to be very effective, retrieving 60% of the game related tweets with a false positive rate (percentage of irrelevant tweets with these keywords) below 5% [20].

#### 3.2 Ground Truth of Event Timing

To evaluate how quickly SportSense can detect an event, we must know the ground truth of the timing of the event, which is surprisingly not readily available. While many media sources (e.g. ESPN, NFL Network) report game events, they only offer game clock timestamps of events. However, because the game clock pauses and resumes during a game, we cannot use a game clock to determine the passage of time between events.

Instead, we observed the broadcasts of 18 games, spanning the 2011 Super Bowl, some 2011-12 regular season games and the 2012 Super Bowl, which include 100 touchdowns. We monitored regular season games at different times and cities, making our sample as random as possible. We manually marked the real clock time of game events. (Due to copyright issues of rebroadcasting games, we were not able to crowdsource this task.)

### 4. Characterizing the Human Response to Events

In this section, we examine the properties of tweets from various users after an event. We analyze the tweet responses of 100 touchdown events across 18 NFL games, investigating characteristics such as the device type, location, user activeness, and the number of words in a tweet. By modeling the nature of the tweets from these groups, we can later develop a precise event detection method (in Section 5).

To analyze this, we use a *Human Delay* metric. Numerically, this is the interval between the event occurrence and the tweet's timestamp. Human delay is influenced by how fast users perceive an event, react to an event, and type a tweet.

We show the tweet response following an event in Figure 1. The x-axis represents the human delay of the tweets, while the y-axis represents the volume of tweets posted per second. We use the error bar to present one standard deviation of uncertainty to represent the variation between the events. From the figure, we can see there are few tweets in the first 50 seconds after an event happens, but the post rates both increase to a peak at 75 seconds after the event happens. Interestingly, among all the games we observed, the first touchdown-related tweet arrived between 7 seconds and 50 seconds after events.

#### 4.1 Mobile Devices Have Higher Post Rate

Many Twitter users post tweets from their mobile devices, e.g., smartphones and tablets. Because of the immediate availability of mobile devices and the significantly different user interfaces, we expect tweets from these different classes of devices to have different delay characteristics.

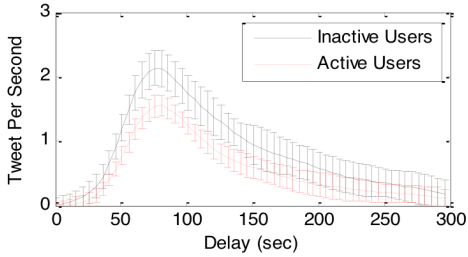

**Figure 3: The tweet responses for active and inactive users.**

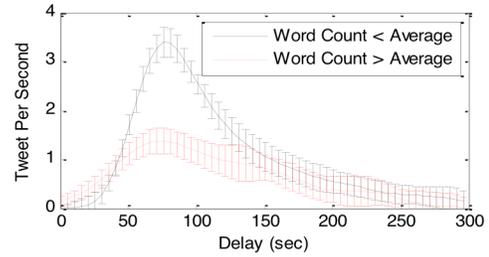

**Figure 4: The tweet responses of short and long tweets.**

We compare mobile and non-mobile responses as shown in Figure 2. The response magnitude on mobile device is twice as large. The peak delays are close; mobile devices reached the peak 2 seconds earlier, but their average delay was longer by 4 seconds. More importantly, we find that the event response on non-mobile devices is more consistent, as evident by the smaller deviation.

To identify tweets posted from mobile devices, we examined the client source information enclosed in the tweet metadata. Approximately 40% of the game-related tweets are posted from Twitter clients on recognizable mobile devices, i.e., iPhone, BlackBerry, Android, HTC, MOTO, as well as mobile browsers and short messages (txt). Tweets on non-mobile clients comprise 30% of the game-related tweets. The remaining 30% tweets are posted from clients that exist in both non-mobile and mobile form. Thus, we do not consider them here.

We hypothesize that typing speed and device/network delay influence the different delays. The typing speed on mobile devices is still lower than that on PC or laptops. Often, users need to unlock the device and open the Twitter application before tweeting. In addition, mobile devices may suffer from slow network conditions [21]. However, mobile devices are more portable and convenient to use when in front of a TV, at home, in bars, etc.

### 4.2 Active Users Post Slower

During NFL games, we find that users have diversified activeness. While most users post fewer than 5 tweets in a game, some post dozens of tweets. We define the activeness as how many tweets the user has posted since the beginning of the game. During the game, we track how many tweets a user has posted and compute the average number of tweets per user. At a certain time, if the user has posted more tweets than the average, they are considered active users. Otherwise, they are considered inactive users.

Next we compare the tweet responses of active and inactive users and illustrate it in Figure 3. For both active and inactive users, the post rates start increasing rapidly from 30 seconds after event happens and reach the peak at the same time. They have similar errors between the template signal and actual signals, but the errors increase when the post rate decreases. Overall, the average and median delays for active users are about 5 seconds longer than for inactive users. Thus, although active users post more frequently than inactive users, their responses are typically more delayed.

### 4.3 More Short Tweets after an Event

In this section we focus on the word count of tweets after an event happens. We observed that users often post short tweets with exclamation marks and repeated letters to express their excitement. e.g., *New York Jets Touchdownnnnn!!!* Therefore, we expect tweets with few words to correlate with the event. In our data, we found that the tweet word count decreased shortly after an event. In particular, the average word count of tweets decreases from 12 words to 6 words within 60 seconds after an event.

To leverage this result, we categorize tweets into long and short tweets by comparing the tweet word count with the average word count value during a game. The average value fluctuates, as tweets are collected in real-time. If a tweet contains more words than the average word count, it will be considered as long tweet. Otherwise, it will be considered as short tweet. Figure 4 shows the tweet response based on tweet word count. The average and median delay of long tweets is 3 and 10 seconds more than short tweets, respectively. More significantly, the post rate of short tweets is twice that of the long tweets.

In addition to the device, activeness, and tweet length, we also attempted to analyze tweets posted from verified users and users in stadiums. However, due to data unavailability (verified and in-stadium users are less than 0.5% of total users), we could not obtain representative results.

### 5. Event Detection

A key technical contribution of the paper is to detect events, such as touchdowns and field goals, from our streams of game-related tweets. SportSense utilizes matched filtering to detect game events. To improve the fidelity of the detector, we create a separate template for different groups of Twitter users, and perform the matched filter separately for each group's tweets. We combine the results of all of the matched filters to achieve accurate and timely event detection. We also elaborate on the proper window size and threshold parameters to achieve the best tradeoff between detection delay and true positive rate.

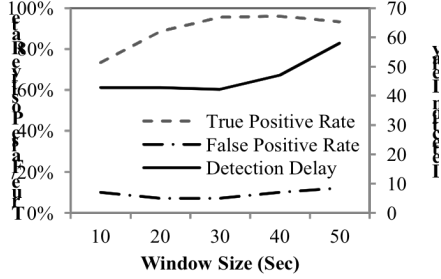

Figure 5: The tradeoff between the event detection delay and the true positive rate.

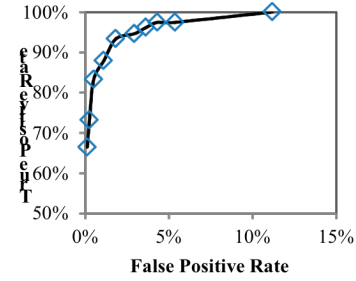

Figure 6: Tradeoff between true positive rate and false positive rate in threshold selection

## 5.1 Event Detection with Matched Filtering

Our event detection strives to analyze Twitter streams to detect events. An event occurrence is routinely followed by a patterned volume of event-related tweets as time progresses. In Section 3.1, we consider tweets that contain a specific event keyword or its variant, e.g., "touchdown" and "TD", as event-related tweets. By aggregating a history of previous events, we can determine the shape of the pattern of tweets that follows an event, i.e., the *event template*. For our event template, we use the tweet response (Figure 1). Our event template accounts for the tweet volume until a certain amount of time has passed after an event. We call this amount of time the *window size* of our event template. To improve the fidelity of the detector, we create a separate template for the different groups of Twitter users that were explored in Section 4.

Once the event templates are established, to perform detection, we adopt *matched filtering*, a technique widely used in signal detection theory to correlate a received signal with a given template. Matched filtering is proven to be the optimal linear filter when dealing with dealing with independent additive noise, and is known to perform well in other scenarios as well [22], The matched filter is applied separately for each user group. The matched filtering process is as follows:

(1) Take our event template **V** over the window **W** and take its time-reversed conjugate to form our matched filter **H**.

$$H(t) = V^*(W - t)$$

(2) Convolve the current window of the ongoing tweet stream **X** with the matched filter **H**.

$$m = X * H$$

(3) If the resulting signal **m** rises above a threshold, our system detects an event occurred at that time.

In our dataset, the interval between two adjacent events is always larger than 5 minutes. If **m** remains above the threshold within 5 minutes after a detected event, we consider it is the same event to prevent repetitive reports of the same event.

## 5.2 Addressing User Diversity

As noted in section 4, we found that tweets from different user categories had different properties. In particular, mobile and inactive users typically exhibit less of a delay of tweeting about an event than non-mobile and active users. Furthermore, users usually post shorter tweets with event keyword after event happened. Noting these characteristic differences, we form different event templates from their unique tweet responses. Thus, the event templates can be seen in Figure 2, Figure 3, and Figure 4

In order to combine the outputs of the various matched filters, $m_i(t)$, to form the output, $m_{comb}(t)$, we evaluated three prominent combination mechanisms; the Maximum Rule, the Mean Rule, and the Product Rule. The *Maximum Rule* calculates the outcome as the maximum of the matched filters. Formally,

$$m_{max}(t) = \max \{ m_i(t) \}$$

The Mean Rule calculates the outcome as the average of the three matched filters. Formally,

$$m_{mean}(t) = 1/3 \cdot \Sigma\{m_i(t)\}$$

The Product Rule calculates the outcome as the product of the three matched filters. Formally,

$$m_{prod}(t) = \Pi\{m_i(t)\}$$

In addition to these three prominent mechanisms, we also evaluated the combination of inactive users and short tweets which exhibit shorter delay.

$$m_{delay}(t) = 1/2 \cdot ( m_{inactive}(t) + m_{short}(t) )$$

We compared the performance of these four methods, and found that the Mean Rule performs best. Indeed, the Mean Rule is known to be especially resilient to noise [23], and useful when inputs are highly correlated. We use $m_{mean}(t)$ in the remainder of the paper, and define $m(t) = m_{mean}(t)$

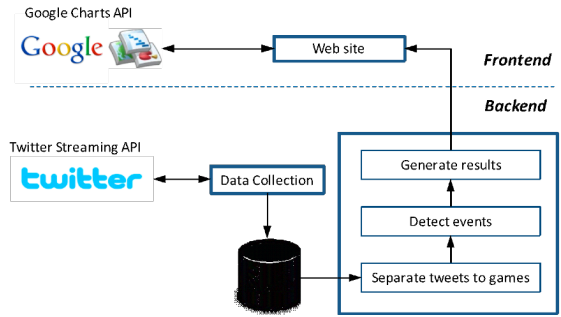

Figure 7: Architecture of SportSense web service implementation.

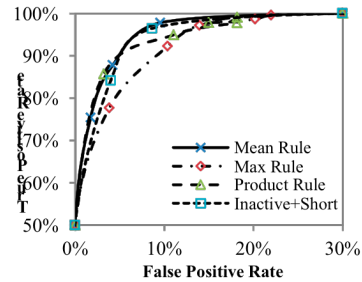

Figure 8: The RoC curves for different combination mechanisms of matched filter detectors. The mean rule outperforms the others.

### 5.3 Parameter Selection

To obtain the optimal window size and threshold, we iterate this matched filter detector method with different window sizes and thresholds through our history of events.

An appropriate window size should be chosen to reduce detection delay while maintaining high accuracy. A short window will not allow for robust event detection; a lengthy time-series is needed to form a pattern for an event template, and the window should be no shorter than this pattern. However, as the window size increases, the delay of event detection also increases, as we need to sample more tweets before a decision can be made about a particular time. Figure 5 shows the effects of window-size on the accuracy and delay of touchdown detection.

Likewise, the choice of threshold parameter impacts the accuracy of the event detection. A low threshold will recognize events when there are none (false negatives), while a high threshold will fail to recognize events (false positives). The tradeoff in threshold choice is shown in Figure 6. An appropriate threshold must be found to balance this tradeoff to fulfill accuracy requirements.

We rely on historical event data to perform optimal parameter selection. With our Touchdown data, we find the optimal window size to be 30 and threshold to be 8. With these parameters, we achieve 97% true positive rate and less than 4% false positives rate, which confirms the efficacy of our Touchdown event detection.

### 6. Implementation and Evaluation

We have implemented SportSense as a real-time web service that visualizes event detection. In this section, we present the details of our web service followed by the thorough evaluation of the performance of SportSense. Finally, we present SportEPG, a sample EPG application built on SportSense. We present the results based on the 2010-11 and 2011-12 NFL seasons.

### 6.1 Web Service Realization

We have implemented the SportSense web service in PHP. It consists of the backend for data collection and analysis, and the frontend for web visualization, as illustrated in Figure 7. We host the web service on Amazon EC2, a reliable and reasonably priced cloud computing platform.

The backend consists of a MySQL database and two modules. The *data collection module* collects game-related tweets through the Twitter Streaming API. The *event detection module* retrieves tweets from the database, assigns them to unique games, and detects game events. The frontend retrieves the results from the event detection module, and visualizes them. We implemented the website frontend using the JavaScript drawing library and Google Charts API. SportSense also provides an API to enable others to leverage analysis results for novel applications. The API provides detected events with timestamps, the temporal trend of posting volume, and the ranking of the most discussed game according to the tweets volume.

The SportSense website has been running for 2 years monitoring NFL games in 2010-11 and 2011-12 Seasons. The event detection method has also been evolving during this period. The original SportSense utilizes the temperature-based method to detect peaks in tweet volume which is similar to existing methods in the related work [1, 12]. However, because our new goal is to achieve event detection within seconds, we devised variable window sizes to reduce detection delay. SportSense starts from the shortest window size and measures (*i*) the percentage increase of the volumes in a sliding window (*ii*) the post volume in the window. If the percentage increase of post volume is high enough and there are enough event-indicative tweets to make a reliable decision, it declares an event has happened. Otherwise, SportSense increments the window size until an event is detected or the largest window size is reached. This method can detect events around 40 seconds after an event happens, with 90% true positive rate and 8% false positive rate [20].

### 6.2 Evaluation

To evaluate the effectiveness of the matched filter detector of SportSense, we use leave-one-out cross-validation (LOOCV) on all the 18 games that we have ground truth as introduced in Section 3.2. LOOCV uses a single game as the validation data, and the remaining games as the training data. There are in total 100 touchdowns, an average of 5 or 6 touchdowns per game. We show that the combination of matched filter detector of different user types can achieve 98% true positive rate and 9% false positive rate.

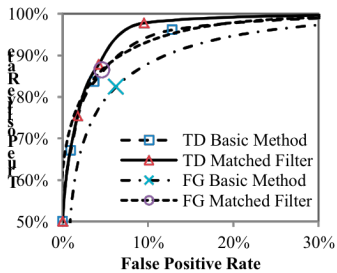

Figure 9: RoC curves for basic temperature based and matched filter detectors. Matched filter performs better for touchdowns and field goals.

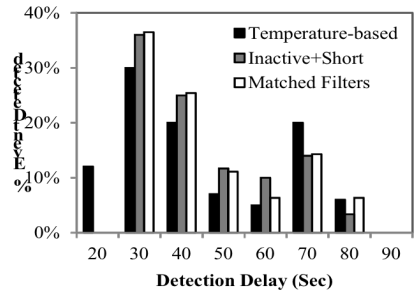

Figure 10: The distribution of event detection delay.

Figure 8 shows the Receiver Operating Characteristics (RoC) Curves for our single matched filter detector and the combination of matched filter detectors for different user groups. The single matched filter detector approach achieves 96% true positive rate and has 13% false positive rate. On the other hand, the combination matched filter detectors performs better, achieving 98% true positive rate and has 9% false positive rate. Figure 9 shows the performance of the basic temperature based method and matched filter method for touchdowns and field goals. The basic temperature-based method performs significantly poorer than the matched filter methods.

We also examine the delay of our detection, which is introduced from two parts, our system and the detection stage. To minimize the system delay, we use parallel data collection and data analysis in our implementation. The delay from the detection stage includes computation delay and the delay due to the matched filter window size. The computation as introduced in Section 5.1 is mainly a convolution operation, which is very fast. Figure 10 shows the delay of events detected. All three methods have an average delay around 45 seconds. The median value of the temperature-based method is 5 seconds shorter than the other two methods but the temperature-based method detects fewer events due to the tradeoff between accuracy and window size. Interestingly, the combination of inactive and short tweets does not outperform the matched filter method. Both have similar detection delay distribution. For the purpose of comparison, we note that EPSN takes about 90 seconds to update their webpage. SportSense detects 60% of events within 40 seconds which is mainly because of the window size. All detected events have delay less than 90 seconds. As a result, SportSense can update results faster than ESPN updates their web page.

### 6.3 SportSense Powered EPG Application

We have developed the SportEPG application to demonstrate the effectiveness of the SportSense system and its APIs. A primary function of electronic program guides, or EPGs, is to assist users in selecting their desired channel. SportEPG achieves this goal by using the SportSense API to identify and highlight the most tweeted programs in real time. Thanks to the flexibility of the keyword-based data collection method, we feed TV program names to the Streaming API to support general TV content. Our event detection is still focused on NFL games. When events take place in sports game, an event icon will show up in the corresponding grid. We leverage AJAX (Asynchronous JavaScript And XML) methods to retrieve analysis results from the SportSense server asynchronously, without interfering with the existing page.

One of the benefits of a real-time updating EPG is that it is accessed frequently and presumably in most cases on a companion device such as a smartphone or a tablet. In turn, the frequent, constant attention to the EPG can be exploited through relevant advertising.

### 7. Conclusion

In this work, we pushed the limit of Twitter as a real-time human sensor. We demonstrated that moderately frequent and diverse social and physical events like those of NFL games can be reliably recognized within 90 seconds using tweets properly collected in real-time. Our results also suggest that the keyword-based method is effective in Twitter-based event detection and the Twitter API limits can be overcome using proper scoping and event detection keywords. We hope this new capability will inspire more applications in the pervasive computing community.

On the other hand, our work is limited to events for which keywords can be predetermined, such as for NFL games. What would be more useful and challenging is to recognize events that are not anticipated, and therefore do not have predefined keywords. Leveraging the work reported here, we are actively pursuing this goal.